\documentclass[12pt]{article}

\renewcommand{\d}{{\rm d}}

\newcommand{\M}{{\cal M}}
\newcommand{\R}{{\cal R}}
\newcommand{\N}{{\cal N}}
\newcommand{\D}{{\cal D}}

\begin{document}

\title{Affine connection form of Regge calculus
}

\author{V.M. Khatsymovsky \\
 {\em Budker Institute of Nuclear Physics} \\ {\em of Siberian Branch Russian Academy of Sciences} \\ {\em
 Novosibirsk,
 630090,
 Russia}
\\ {\em E-mail address: khatsym@gmail.com}}
\date{}
\maketitle
\begin{abstract}
Regge action is represented analogously to how the Palatini action for general relativity (GR) as some functional of the metric and a general connection as independent variables represents the Einstein-Hilbert action.

The piecewise flat (or simplicial) spacetime of Regge calculus is equipped with some world coordinates and some piecewise affine metric which is completely defined by the set of edge lengths and the world coordinates of the vertices. The conjugate variables are the general nondegenerate matrices on the 3-simplices which play a role of a general discrete connection. Our previous result on some representation of the Regge calculus action in terms of the local Euclidean (Minkowsky) frame vectors and orthogonal connection matrices as independent variables is somewhat modified for the considered case of the general linear group GL(4,R) of the connection matrices.

As a result, we have some action invariant w. r. t. arbitrary change of coordinates of the vertices (and related GL(4,R) transformations in the 4-simplices). Excluding GL(4,R) connection from this action via the equations of motion we have exactly the Regge action for the considered spacetime.

\end{abstract}


MSC classes: 83C27; 53C05

keywords: Einstein theory of gravity; minisuperspace model; piecewise flat spacetime; Regge calculus; affine connection; Palatini action; discrete connection

Minisuperspace models allow one to work with a countable number of the degrees of freedom. This may be important in such an essentially nonlinear theory of gravity as GR. In quantum framework, minisuperspace approach works as a kind of lattice methods \cite{Ham'} allowing to solve the problem of the formal nonrenormalizability of the original continuum GR. The GR considered as some curved geometry theory can be most naturally formulated on a minisuperspace by restricting ourselves by the metric field distributions $g_{\lambda \mu} (x )$ describing the piecewise flat spacetimes composed of the flat 4D tetrahedra ({\it 4-simplices}) or the {\it simplicial complex} \cite{piecewiseflat=simplicial'}. These spacetimes can be chosen arbitrarily close in some sense to any given Riemannian spacetime, and GR on them is known as Regge calculus \cite{Regge'}; see, e. g. review in Ref. \cite{RegWil'}. The Regge action ($\int R \sqrt{g} \d^4 x$) is
\begin{equation}                                                            
\label{S-Regge} 2 \sum_{\sigma^2}{\alpha (\sigma^2) A (\sigma^2)},
\end{equation}

\noindent where $A (\sigma^2)$ is the area of the triangle (2-simplex) $\sigma^2$, $\alpha (\sigma^2)$ is the angle defect on this triangle, summation is over all the 2-simplices $\sigma^2$. The action and the geometry of the piecewise flat spacetime is completely defined by the full set of the edge lengths. Among contemporary approaches the Causal Dynamical Triangulations approach related to the Regge calculus has led to important results in quantum gravity \cite{cdt}.

The Regge action and the corresponding equations of motion in terms of edge lengths seem to be simple from geometrical point of view, but rather complicated from the field theoretical point of view. In the continuum GR, certain simplification is achieved by adding some extra independent variables so that some new action in terms of these is classically (on the equations of motion) equivalent to the original Einstein-Hilbert action, but is of the first order contrary to the second order Einstein-Hilbert one. The discrete analogs of these first order formulations have been widely addressed. The discrete analogs of the connection and curvature, finite orthogonal rotations, were first considered in \cite{Fro}. An application to the discrete Hamiltonian analysis of gravity was discussed in \cite{Ban3}. In the paper \cite{our2}, we have found the exact representation of the Regge action using the discrete analogs of the connection and curvature \cite{Fro}. An approximate version was considered in \cite{CasDadMag}. A first order form of Regge calculus was considered in \cite{Bar} where the extra independent variables are the interior dihedral angles of a simplex, with conjugate variables the areas of the triangles.

Now we consider the possibility of using general linear rather than the orthogonal matrices as discrete connection variables. Thus, we are talking about some discrete version of the Palatini action \cite{Pal} depending on the metric and a general connection as independent variables. The Palatini action can be written in the form
\begin{eqnarray}\label{Pal action}                                          
& & \hspace{0mm} S = \int R^\nu_{\lambda \nu \mu} g^{\lambda \mu} \sqrt {g} \d^4 x = \int [ - \Gamma^\nu_{\lambda \mu} \partial_\nu (\sqrt{g} g^{\lambda \mu}) + \Gamma^\nu_{\lambda \nu} \partial_\mu (\sqrt{g} g^{\lambda \mu}) \nonumber \\ & & + \sqrt{g} g^{\lambda \mu} ( \Gamma^\nu_{\lambda \mu} \Gamma^\rho_{\nu \rho} - \Gamma^\nu_{\lambda \rho} \Gamma^\rho_{\nu \mu} ) ]  \d^4 x,
\end{eqnarray}

\noindent where $g \equiv \det \| g_{\lambda \mu} \|$ and we assume the Euclidean metric signature for definiteness. Here, the metric $g_{\lambda \mu}$ and the Christoffel symbol $\Gamma^\lambda_{\mu \nu}$ are independent variables, and an additional requirement for that the equations of motion would give for $\Gamma^\lambda_{\mu \nu}$ the unique metric-compatible connection for $g_{\lambda \mu}$ is the requirement that the connection be torsion-free, $\Gamma^\lambda_{\mu \nu} = \Gamma^\lambda_{\nu \mu}$. Note, however, that $\Gamma^\lambda_{\mu \nu} \d x^\nu$ describes an infinitesimal transformation matrix of any vector being parallel transported from $x^\lambda$ to $x^\lambda + \d x^\lambda$. Some discrete analog of it would be a finite transformation matrix $\M^\lambda_{\sigma^3 \mu}$ for the transport across the 3-simplex $\sigma^3$, the 3-dimensional face between some two neighboring 4-simplices. Consequently, imposing the zero torsion condition require comparing the matrices $\M^\lambda_{\sigma^3 \mu}$ at the different $\sigma^3$s, that is, this condition is not local and is unnatural in the considered formalism.

Fortunately, the requirement that the connection be torsion-free is not absolutely necessary for the GR action being reproduced via equations of motion in the Palatini formalism. Without this condition, we get for $\Gamma^\lambda_{\mu \nu}$ the unique metric-compatible part $\Gamma^\lambda_{\mu \nu} ( \{ g_{\lambda \mu} \} )$ plus some part $\Gamma_\nu \delta^\lambda_\mu $ with torsion which, however, does not contribute to the action.

Thus, it is more natural to consider the completely unrestricted three-index variable $\Gamma^\lambda_{\mu \nu}$ as the continuum counterpart; the discrete counterpart will be the general nondegenerate transformation matrices $4 \times 4$ on the 3-simplices $\sigma^3$, $\M_{\sigma^3} \in $ GL(4, R).

If the connection has support on the 3-simplices, this means that the metric is piecewise constant or constant in the 4-simplices. That is, the world coordinate frame is piecewise affine. The metric field $g_{\lambda \mu} (x )$ is completely defined by the edge lengths and the coordinates of the simplicial vertices $\sigma^0$. Earlier we have found a representation of the Regge action in terms of the edge vectors and orthogonal connection matrices between the  Euclidean local frames in the 4-simplices \cite{our2}. Now we should, first, transform this expression from the local metric $diag (1, 1, 1, 1)$ to the general symmetric nondegenerate constant matrix $g_{\lambda \mu}$ in each 4-simplex. Second, we should generalize it from the orthogonal to general nondegenerate connection matrices.

The considered expression for the Regge action connection representation in the local Euclidean frame formalism takes the form
\begin{eqnarray}\label{Cartan-SO(4)}                                        
\hspace{0mm} S^{\rm discr}_{\rm SO(4)} = 2 \sum_{\sigma^2} A(\sigma^2 ) \arcsin \left [ \frac{ R^{ab}_{\sigma^2} ( \Omega ) l^c_{\sigma^1_1} l^d_{\sigma^1_2} }{4 A(\sigma^2 )} \epsilon_{abcd} \right ].
\end{eqnarray}

\noindent Here, the edge vectors $l^a_{\sigma^1_1}$, $l^a_{\sigma^1_2}$ form the triangle $\sigma^2$, the area of $\sigma^2$ is $A(\sigma^2 )$ $=$ \\ $\sqrt{ l^2_{\sigma^1_1} l^2_{\sigma^1_2} - (l_{\sigma^1_1} l_{\sigma^1_2})^2} / 2 $. The {\it curvature} SO(4) matrix $R^{ab}_{\sigma^2} ( \Omega )$ on the triangles $\sigma^2$ is the product of the connection SO(4) matrices $\Omega_{\sigma^3}$s for the set of $\sigma^3$s meeting at $\sigma^2$ ordered along a closed path encircling $\sigma^2$ and passing through each of these (and only these) $\sigma^3$s,
\begin{equation}                                                            
R_{\sigma^2} (\Omega ) = \prod_{ \{ \sigma^3 : ~ \sigma^3\supset\sigma^2 \} }{\Omega^{\epsilon (\sigma^2, \sigma^3)}_{\sigma^3}},
\end{equation}

\noindent where $\epsilon (\sigma^2, \sigma^3) = \pm 1$ is some sign function. This path begins and ends in a 4-simplex $\sigma^4$. That is, $R^{ab}_{\sigma^2}$ is defined in (the frame of) this simplex. The edge vectors $l^a_{\sigma^1_1}$, $l^a_{\sigma^1_2}$ are also defined in this simplex.

When we pass from the local Euclidean frames to the piecewise affine world one, the edge vectors $l^a$ are substituted by the corresponding world coordinate differences $\Delta x^\lambda$. The RHS of (\ref{Cartan-SO(4)}) takes the form
\begin{eqnarray}\label{Cartan-SO(4)-version}                                
\hspace{0mm} 2 \sum_{\sigma^2} A(\sigma^2 ) \arcsin \left [ \frac{ \R^\lambda_{\sigma^2 \tau} ( \M ) g^{\tau \mu} \Delta x^\nu_{\sigma^1_1} \Delta x^\rho_{\sigma^1_2} }{4 A(\sigma^2 )} \epsilon_{\lambda \mu \nu \rho} \sqrt{g} \right ]
\end{eqnarray}

\noindent where $\R_{\sigma^2} ( \M )$ is built of $\M_{\sigma^3}$s just as $R_{\sigma^2} ( \Omega )$ is built of $\Omega_{\sigma^3}$s. The generalization to arbitrary $\M_{\sigma^3}$s should provide the correct equations of motion. The dependence of $\R_{\sigma^2}$ on $\M_{\sigma^3}$ takes the form $(\Gamma_1 (\sigma^2 , \sigma^3 ) \M_{\sigma^3} \Gamma_2 (\sigma^2 , \sigma^3 ))^{\epsilon (\sigma^2, \sigma^3)}$, $\sigma^2 \subset \sigma^3$, $\Gamma_1 (\sigma^2 , \sigma^3 )$, $\Gamma_2 (\sigma^2 , \sigma^3 )$ are GL(4, R) matrices. Let us take for example $\epsilon (\sigma^2, \sigma^3) = + 1$ (that is, $\R_{\sigma^2}$ is linear in $\M_{\sigma^3}$). Applying $\M^\nu_{\sigma^3 \lambda} \partial / \partial \M^\nu_{\sigma^3 \mu}$ to (\ref{Cartan-SO(4)-version}) gives
\begin{eqnarray}\label{eq-motion-version}                                   
- \hspace{-10mm} \sum_{ \hspace{10mm} \{\sigma^2 : ~ \sigma^2 \subset \sigma^3 \} } \hspace{-10mm} [ \Gamma_2 (\sigma^2 , \sigma^3 ) \frac{v_{\sigma^2 } \R_{\sigma^2 }}{\cos \alpha (\sigma^2 )} \Gamma^{-1}_2 (\sigma^2 , \sigma^3 ) ]^\mu {}_{ \lambda}, ~~~ v_{\sigma^2 \lambda \mu} = \frac{1}{2} \sqrt{g} \epsilon_{\lambda \mu \nu \rho} \Delta x^\nu_{\sigma^1_1} \Delta x^\rho_{\sigma^1_2},
\end{eqnarray}

\noindent for the contribution to the equations of motion. Here $\alpha (\sigma^2 )$ means just $\arcsin$ function in (\ref{Cartan-SO(4)-version}). For the particular ansatz of the metric-compatible connection with $\R_{\sigma^2}$ rotating around $\sigma^2$ (by the angle $\alpha (\sigma^2 )$),
\begin{equation}\label{ansatz-identity}                                     
v_{\sigma^2 } \R_{\sigma^2 } + \R^{-1}_{\sigma^2 } v_{\sigma^2 } = 2 v_{\sigma^2 } \cos \alpha (\sigma^2 ), ~~~ \R_{\sigma^2 } v_{\sigma^2 } + v_{\sigma^2 } \R^{-1}_{\sigma^2 } = 2 v_{\sigma^2 } \cos \alpha (\sigma^2 ). \end{equation}

\noindent The expression (\ref{eq-motion-version}) looks similar to the sum of the bivectors $v_{\sigma^2}$ of the faces $\sigma^2$ of $\sigma^3$, but not quite that. The situation can be corrected by the following replacement of the matrix describing the curvature in (\ref{Cartan-SO(4)-version}),
\begin{equation}                                                            
\R_{\sigma^2 } \Longrightarrow \frac{1}{2} ( \R_{\sigma^2 } - \R^{-1}_{\sigma^2 } ).
\end{equation}

\noindent Geometrically, the curvature is evaluated by parallel transport of the vector along a closed loop in the two possible opposite directions and comparing the results. This gives a discrete action,
\begin{eqnarray}\label{Palatini}                                            
\hspace{0mm} S^{\rm discr}_{\rm GL(4,R)} = 2 \sum_{\sigma^2} A(\sigma^2 ) \arcsin \left \{ \frac{ [\R_{\sigma^2} - \R^{-1}_{\sigma^2} ]^\lambda {}_{ \tau} ( \M ) g^{\tau \mu} \Delta x^\nu_{\sigma^1_1} \Delta x^\rho_{\sigma^1_2} }{8 A(\sigma^2 )} \epsilon_{\lambda \mu \nu \rho} \sqrt{g} \right \}.
\end{eqnarray}

\noindent Here, to resume,
\begin{equation}                                                           
\R_{\sigma^2} (\M ) = \prod_{ \{ \sigma^3 : ~ \sigma^3\supset\sigma^2 \} }{\M^{\epsilon (\sigma^2 \sigma^3)}_{\sigma^3}}
\end{equation}

\noindent  ($\epsilon (\sigma^2, \sigma^3) = \pm 1$ is some sign function), the product of the connection GL(4,R) matrices $\M_{\sigma^3}$s for the set of $\sigma^3$s meeting at $\sigma^2$ ordered along a closed path encircling $\sigma^2$ and passing through each of these (and only these) $\sigma^3$s. This path begins and ends in a 4-simplex $\sigma^4$, and the metric appearing in (\ref{Palatini}) and the dual bivector $v_{\sigma^2 \lambda \mu}$ are taken just in this 4-simplex. Again, denote $\R_{\sigma^2} = (\Gamma_1 (\sigma^2 , \sigma^3 ) \M_{\sigma^3} \Gamma_2 (\sigma^2 , \sigma^3 ))^{\epsilon (\sigma^2, \sigma^3)}$, apply $\M^\nu_{\sigma^3 \lambda} \partial / \partial \M^\nu_{\sigma^3 \mu}$ to the action and take into account the following equality for differentiating the inverse matrix,
\begin{equation}                                                           
\M^\nu {}_\lambda \frac{\partial }{\partial \M^\nu {}_\mu} (\M^{-1})^\rho {}_\tau = - (\M^{-1})^\mu {}_\tau \delta^\rho_\lambda.
\end{equation}

\noindent The equations of motion take the form
\begin{eqnarray}\label{eq-motion}                                          
\hspace{-10mm} \sum_{ \hspace{10mm} \{\sigma^2 : ~ \sigma^2 \subset \sigma^3 \} } \hspace{-10mm} \epsilon (\sigma^2, \sigma^3) \left [ \Gamma_2 (\sigma^2 , \sigma^3 ) \frac{v_{\sigma^2 } \R^{\epsilon (\sigma^2, \sigma^3)}_{\sigma^2 } + \R^{- \epsilon (\sigma^2, \sigma^3)}_{\sigma^2 } v_{\sigma^2 } }{\cos \alpha (\sigma^2 )} \Gamma^{-1}_2 (\sigma^2 , \sigma^3 ) \right ] = 0.
\end{eqnarray}

\noindent Now, with taking into account (\ref{ansatz-identity}) for the particular ansatz of the metric-compatible connection, these equations express the closure of the (dual) bivectors \\ $\Gamma_2 (\sigma^2 , \sigma^3 ) v_{\sigma^2} \Gamma^{-1}_2 (\sigma^2 , \sigma^3 )$, $\sigma^2 \subset \sigma^3$ fulfilled identically; the matrices $\Gamma_2$ serve to transport them all to the same 4-simplex, one of the two containing the given $\sigma^3$ (from which $\M_{\sigma^3}$ transports).

Thus, the $S^{\rm discr}_{\rm GL(4,R)}$ (\ref{Palatini}) can serve as an exact affine connection representation of the Regge action. The configuration of the considered system is completely described by the set of connection matrices on the 3-simplices $\M_{\sigma^3} \in {\rm GL(4,R)}$, edge lengths $l_{\sigma^1}$ and coordinates of the vertices $x^\lambda_{\sigma^0}$. The coordinates of the vertices $x^\lambda_{\sigma^0}$ define the contravariant edge vector
\begin{equation}                                                           
\Delta x^\lambda_{\sigma^1} = x^\lambda_{\sigma^0_2 } - x^\lambda_{\sigma^0_1 }
\end{equation}

\noindent for the edge $\sigma^1$, the difference between the coordinates of its ending vertices $\sigma^0_1$, $\sigma^0_2$. Knowing the contravariant edge vectors and edge lengths allows to find the metric tensor in each 4-simplex as a function of its ten edge lengths from simple system of ten linear equations,
\begin{equation}                                                           
\Delta x^\lambda_{\sigma^1} \Delta x^\mu_{\sigma^1} g^{\lambda \mu } = l^2_{\sigma^1 }.
\end{equation}

Any gauge transformation is an arbitrary change of coordinates of the vertices. This is analogous to diffeomorphisms in the continuum theory. Elementary gauge transformation is an arbitrary change of coordinates of any {\it given} vertex $\sigma^0$. When doing so, the vectors $\Delta x^\lambda_{\sigma^1 }$ are changed for the edges containing this vertex, $\sigma^1 \supset \sigma^0$. This can be described as general coordinate transformations acting in each 4-simplex containing the vertex, $\sigma^4 \supset \sigma^0$,
\begin{equation}                                                           
\Delta x^\lambda_{\sigma^1} \Longrightarrow \Delta \tilde{x}^\lambda_{\sigma^1} = \N^\lambda_{\sigma^4 \mu} \Delta x^\mu_{\sigma^1}, ~~~ \sigma^4 \supset \sigma^1
\end{equation}

\noindent and
\begin{equation}                                                           
A_\lambda \Longrightarrow \tilde{A}_\lambda = A_\mu (\N^{-1}_{\sigma^4})^\mu {}_\lambda
\end{equation}

\noindent for the covariant vectors in $\sigma^4$. This $\N_{\sigma^4}$ is restricted by the condition that it does not change the vectors of those edges of $\sigma^4$ which do not contain $\sigma^0$. Besides that, the matrices in the neighboring 4-simplices $\N_{\sigma^4_1}$ and $\N_{\sigma^4_2}$ should give the same result when acting on the vectors of their common edges $\sigma^1 \subset \sigma^4_1 \cap \sigma^4_2$. $S^{\rm discr}_{\rm GL(4,R)}$ is invariant w. r. t. such transformations:
\begin{equation}                                                           
\M^\lambda_{\sigma^3 \mu} \Longrightarrow \tilde{\M }^\lambda_{\sigma^3 \mu} = \N^\lambda_{\sigma^4_2 \nu} \M^\nu_{\sigma^3 \rho} (\N^{-1}_{\sigma^4_1})^\rho {}_\mu, ~~~ \sigma^3 = \sigma^4_1 \cap \sigma^4_2
\end{equation}

\noindent ($\M_{\sigma^3}$ acts from $\sigma^4_1$ to $\sigma^4_2$), the curvature matrix
\begin{equation}                                                           
\R^\lambda_{\sigma^2 \mu} \Longrightarrow \tilde{\R }^\lambda_{\sigma^2 \mu} = \N^\lambda_{\sigma^4 \nu} \R^\nu_{\sigma^2 \rho} (\N^{-1}_{\sigma^4})^\rho {}_\mu,
\end{equation}

\noindent $\sigma^4$ is the considered above 4-simplex where $\R_{\sigma^2}$ is defined, and (\ref{Palatini}) is formally invariant w. r. t. any $\N_{\sigma^4}$ acting in this $\sigma^4$.

The invariance under the general transformation of the vertex coordinates says that these coordinates might be chosen by hand in some convenient way. In any given 4-simplex this can be done but procedure of a convenient extension to the other 4-simplices is unknown in general case. In some particular cases of simple structure this choice can be made globally. For example, there is the case of the periodic structure such that spacetime is decomposed into 4-cubes and each cube is decomposed with the help of its edges and diagonals into the 4-simplices \cite{RocWil}. Here we have, in fact, the 4-cubic lattice, the vertices of which can be numbered by integer valued Cartesian coordinates. Thereby, the action can be explicitly written in terms of the edge lengths as metric variables (and arbitrary nondegenerate connection matrices, of course).

If the considered affine connection action is compared with its orthogonal connection form, the latter differs by fewer number of the connection degrees of freedom (6 parameters of the general orthogonal matrix instead of 16 elements of general 4 $\times$ 4 matrix) although at the expense of imposing additional (orthogonality) constraints, and by considerably larger tetrad/metric sector with a number of additional constraints. For example, consider the simplicial periodic structure \cite{RocWil}. There are 24 4-simplices in each 4-cube and we have some 16-component tetrad of edge vectors in the local frame of each 4-simplex $l^a_{\sigma^1_i | \sigma^4}, i = 1, 2, 3, 4$. Simultaneously, a number of some constraints should be imposed of the type of $(l^a_{\sigma^1 | \sigma^4_1})^2 = (l^a_{\sigma^1 | \sigma^4_2})^2$ for the vectors of the same edge $\sigma^1$ defined in the different local frames of some two 4-simplices sharing this edge, $\sigma^1 \subset \sigma^4_1 \cap \sigma^4_2$. In contrast, in the affine connection formalism, we have the lengths of the 15 edges and 4 coordinates of the vertex per 4-cube. The coordinates of the vertex can be conveniently chosen in the overall lattice as, e. g. integer-valued coordinates, and we are left just with 15 edge lengths per cube/vertex. This may also have an advantage in a numerical simulation.

Certain inconvenience associated with using the affine connection is that the freely chosen elements $\M_{\sigma^3} \in {\rm GL(4,R)}$ do not automatically enter the domain of definition of $S^{\rm discr}_{\rm GL(4,R)}$. This is because the argument of $\arcsin$ in (\ref{Palatini}) can become in absolute value greater than unity if we rescale $\M_{\sigma^3}$s strongly enough. Therefore, we need to specially check whether we came out of the domain of definition of the action.

An attractive feature of the affine connection may be a uniform description with the same group GL(4,R) of both the Riemannian and pseudo-Riemannian spacetimes instead of the two groups SO(4) and SO(3,1) in the local frame formalism.

If the functional integral is considered, some typical value of the coefficient at $\arcsin$ in the functional integral exponent is $A / l^2_g$ where $A$ is a typical simplicial area and $l_g$ is the Plank scale $10^{-33} cm$. Therefore, at $A \gg l^2_g$ the greatest contribution, not suppressed by the oscillating exponential, comes from the region of small argument of $\arcsin$, and $\arcsin$ can be substituted by its argument. We have certain matrix analog of some Bessel function, which is transformed into an absolutely convergent integral by the transition to the contour integration in the complex plane,
\begin{equation}\label{func-int-model}                                     
\int^{\infty}_0 \exp \left [ i\frac{A}{l^2_g} \left ( x - \frac{1}{x} \right ) \right ] \frac{\d x}{x} \Longrightarrow \int^{\infty}_0 \exp \left [ -\frac{A}{l^2_g} \left ( x + \frac{1}{x} \right ) \right ] \frac{\d x}{x},
\end{equation}

\noindent where A stands for a typical scale of the dual bivector $v_{\sigma^2 \lambda}{}^\tau / 2$ (\ref{eq-motion-version}) in (\ref{Palatini}) (now with the standard replacement $\sqrt{g} \to \sqrt{ - g}$ for the pseudo-Riemannian spacetime), that is, just the triangle area. Here we suggest that the functional integral measure includes the product over the 3-simplices $\prod_{\sigma^3} \D \M_{\sigma^3}$ where $\D \M = (\det \M )^{-4} \d^{16} \M $ is the Haar measure on GL(4,R). Using multiplicative dependence of $\R$s on $\M$s and invariance of the Haar measure w. r. t. the multiplication we can make certain change of variables and replace some $\D \M_{\sigma^3}$s by $\D \R_{\sigma^2}$s for those $\R_{\sigma^2}$s that are not connected with each other (by the Bianchi identities \cite{Regge'}). In (\ref{func-int-model}), $\d x / x$ just models $\D \R$. (Of course, the genuine expression is not factorizable into the analogs of (\ref{func-int-model}) so easily.) The above suggests that the contribution of areas exceeding the Plank scale is probably exponentially suppressed. Recasting to an absolutely convergent integral may also be important for a possible numerical simulation.

In conclusion, using the affine connection form of Regge calculus may have interesting implications in both analytical and numerical analysis of classical and especially quantum GR.

\section*{Acknowledgments}

The present work was supported by the Ministry of Education and Science of the Russian Federation


\begin{thebibliography}{99}
\bibitem{Ham'}
 H. W. Hamber, Quantum Gravity on the Lattice, {\it Gen. Rel. Grav.} {\bf 41}, 817 (2009); ({\it Preprint} arXiv:0901.0964[gr-qc]).
\bibitem{piecewiseflat=simplicial'}
 J. Cheeger, W. M\"{u}ller, and R. Shrader,  On the curvature of the piecewise flat spaces, {\it Commun. Math. Phys.} {\bf 92}, 405 (1984).
\bibitem{Regge'}
 T. Regge, General relativity theory without coordinates, {\it Nuovo Cimento} {\bf 19}, 558 (1961).
\bibitem{RegWil'}
 T. Regge and R. M. Williams, Discrete structures in gravity, {\it Journ. Math. Phys.} {\bf 41}, 3964 (2000); ({\it Preprint} arXiv:0012035[gr-qc]).
\bibitem{cdt}
 J. Ambjorn, A. Goerlich, J. Jurkiewicz, and R. Loll,  Nonperturbative Quantum Gravity, {\it Physics Reports} {\bf 519}, 127 (2012); ({\it Preprint} arXiv:1203.3591[hep-th]).
\bibitem{Fro}
 J. Fr\"{o}hlich, Regge Calculus and Discretized Gravitational Functional
 Integrals, IHES preprint, 1981 (unpublished); in {\it Non-Perturbative Quantum Field Theory: Mathematical Aspects and Applications, Selected Papers}, 523 (Singapore: World Scientific, 1992).
\bibitem{Ban3}
 M. Bander, Hamiltonian lattice gravity. II. Discrete moving-frame for\-mu\-la\-tion
 {\it Phys. Rev.} D {\bf 38}, 1056 (1988).
\bibitem{our2}
 V. M. Khatsymovsky, Tetrad and self-dual formulations of Regge calculus, {\it Class. Quantum Grav.} {\bf 6}, L249 (1989).
\bibitem{CasDadMag}
 M. Caselle, A. D'Adda, and L. Magnea, Regge calculus as a local theory of the Poincar\'{e}
 group, {\it Phys. Lett.} B {\bf 232}, 457 (1989).
\bibitem{Bar}
 J. W. Barrett, First order Regge calculus, {\it Class. Quantum Grav.} {\bf 11} 2723 (1994); ({\it Preprint} arXiv:9404124[hep-th]).
\bibitem{Pal}
 A. Palatini, Deduzione invariantiva delle equazioni gravitazionali dal principio di Hamilton, {\it R.C.
 Circ. Mat. Palermo} {\bf 43}, 203 (1919).
\bibitem{RocWil}
 M. Rocek, R.M. Williams, Quantum Regge calculus, {\it Phys. Lett.} B {\bf 104}, 31 (1981).

\end{thebibliography}
\end{document}